\documentclass[useAMS,usenatbib]{mn2e}
\usepackage{amssymb,amsmath}
\usepackage{epsf}
\usepackage{subfigure}
\usepackage{graphicx}

\title[Deriving an X-ray luminosity function]{Deriving an X-ray luminosity function of dwarf novae based on parallax measurements}
\author[K. Byckling, K. Mukai, J.R. Thorstensen, J.~P. Osborne]{K. Byckling$^{1}$\thanks{E-mail:
kjkb2@star.le.ac.uk}
K. Mukai$^{2}$, J.R. Thorstensen$^{3}$ and J.~P. Osborne$^{1}$ \\
$^{1}$Department of Physics and Astronomy, University of Leicester, University Road, Leicester, LE1 7RH, UK\\
$^{2}$NASA/Goddard Space Flight Center, Greenbelt, MD 20771, USA \\
$^{3}$Department of Physics and Astronomy, 6127 Wilder Laboratory, Dartmouth College, Hanover, NH 03755-3528, USA}
\begin{document}

\date{Accepted . Received ; in original form}

\pagerange{\pageref{firstpage}--\pageref{lastpage}} \pubyear{}

\maketitle

\label{firstpage}

\begin{abstract}
We have derived an X-ray luminosity function using parallax-based
distance measurements of a set of 12 dwarf novae, consisting of {\it
Suzaku}, {\it XMM-Newton} and {\it ASCA} observations. The shape of
the X-ray luminosity function obtained is the most accurate to date,
and the luminosities of our sample are concentrated between $\sim$
10$^{30}$--10$^{31}$ erg s$^{-1}$, lower than previous measurements of
X-ray luminosity functions of dwarf novae. Based on the integrated
X-ray luminosity function, the sample becomes more incomplete below
$\sim$ 3 $\times$ 10$^{30}$ erg s$^{-1}$ than it is above this
luminosity limit, and the sample is dominated by X-ray bright dwarf
novae. The total integrated luminosity within a radius of 200~pc is
1.48 $\times$ 10$^{32}$ erg s$^{-1}$ over the luminosity range of 1
$\times$ 10$^{28}$ erg s$^{-1}$ and the maximum luminosity of the
sample (1.50 $\times$ 10$^{32}$ erg s$^{-1}$). The total absolute
lower limit for the normalised luminosity per solar mass is 1.81
$\times$ 10$^{26}$ erg s$^{-1}$ M$_{\odot}^{-1}$ which accounts for
$\sim$ 16 per cent of the total X-ray emissivity of CVs as estimated
by \citet{saz06}.
\end{abstract}


\section{Introduction}
Cataclysmic variables, i.e.~CVs consist of an accreting white dwarf
primary and a late-type main sequence star, and accrete via Roche lobe
overflow. CVs can be divided into several subclasses of which so
called dwarf novae (DNe) are the most numerous subclass of CVs in our
Galaxy. In these systems, the white dwarf has a weak magnetic field
strength \citep[B $\lesssim$ 10$^{6}$ G,][]{van96} compared to
magnetic CVs, such as polars, and thus the formation of an accretion
disc is possible. From time to time, the disc is seen to brighten by
several magnitudes lasting from days to several weeks. This
brightening of the disc, i.e. an outburst, is thought to be caused by
disc instabilities, which are described in detail in \citet{las01}. In
quiescence, DNe are sources of optical emission emanating from the
accretion disc and the bright spot where the material from the
secondary hits the edge of the disc. Quiescent optical spectra of DNe
are characterized by strong Balmer emission lines and weaker He I
lines with some heavier elements. Also, DNe are sources of hard X-rays
which are thought to originate from an optically thin boundary layer
during quiescence. However, during an outburst hard X-rays are
quenched as the boundary layer becomes optically thick and thus a
source of soft X-rays and EUV emission \citep[][]{pri77,pri79}.

At the time of the discovery of the Galactic Ridge X-ray Emission
(GRXE) in 1982 \citep{wor82}, discrete point sources were thought to
be the origin of the GRXE emission. However, the origin has been
debated since, but observational evidence gathered to date since the
GRXE discovery supports the view that the GRXE is not due to diffuse
origin but due to discrete point sources, such as CVs and other
accreting binary systems \citep[see recent studies by
e.g.][]{rev07,rev08}. More supporting evidence was given by the recent
{\it Chandra} study carried out by \citet{rev09} who resolved over 80
per cent of the GRXE into point sources in the 6--7 keV energy range
during an ultra-deep 1 Msec observation. 

Based on {\it EXOSAT} observations of the X-ray emission in the
Galactic Plane, \citet{war85} concluded that if the GRXE is assumed to
be originating from discrete point sources, the bulk of the emission
observed must be due to a population of low luminosity X-ray sources
(L$_{x}$ $<$ 10$^{33.5}$ erg s$^{-1}$), such as CVs. Subsequently,
\citet{muk93} suggested that DNe could significantly contribute to the
GRXE based on their study of an {\it EXOSAT} Medium Energy (ME) DN
sample. According to this study, the space density of DNe is
sufficiently high to account for a significant fraction of the
GRXE. Later on, \citet{ebi01} resolved sources down to 3 $\times$
10$^{-15}$ erg cm$^{-2}$ s$^{-1}$ in their {\it Chandra} observation
of the Galactic Ridge, equivalent to $L_{x}$ $>$ 2.3 $\times$
10$^{31}$ erg s$^{-1}$ at 8~kpc, concluding that the number of
resolved point sources above this level is insufficient for them to be
the major contributor to the GRXE. Since these previous works have not
completely resolved the contribution of CVs to the GRXE, further
studies are needed. As was noted by \citet{muk93}, unbiased and
sensitive surveys with accurate distance measurements of CVs are
needed. This way, accurate X-ray luminosity functions (XLFs) can be
obtained, and the contribution to the GRXE estimated more precisely.

The motivation for our work was mainly given by the inaccuracies in
the XLFs of Galactic CV populations, such as those by \citet{bas05}
and \citet{saz06}. Baskill et al. derived an XLF using 34 {\it ASCA}
observations of non-magnetic CVs (including 23 DN observations). Their
sample lacked accurate distance measurements as only 10 sources had
parallax-based distance measurements. Furthermore, this sample was
biased by high X-ray flux sources since {\it ASCA} was intended to be
a spectroscopic mission and the sources in the studied sample were
known to be X-ray bright. Also, the {\it ASCA} study was purely
archival without any sample selection (e.g. the distance was not
limited) as Baskill et al. chose all non-magnetic CV observations in
the archive, they did not filter out sources which were in an outburst
state, or restrict the study to one type of objects only. The XLF
study by \citet{saz06} focused on building up an XLF in the 2--10~keV
range combining the {\it RXTE} Slew Survey (XSS) and {\it ROSAT}
All-Sky Survey (RASS) observations of active binaries, CVs and young
main sequence stars in the luminosity range $\sim$ 10$^{27.5}$ $<$
$L_{x}$ $<$ 10$^{34}$ erg s$^{-1}$. However, uncertainties in the
luminosities in this study were introduced by inadequate accuracies in
the distances, for example, many of the intermediate polars (IPs) in
their sample had poorly known distances. Only a few sources had
parallax measurements from, e.g., the Hipparcos or Tycho catalogues
(astrometric uncertainties $\sim$ 1 mas) and ground-based parallax
measurements from \citet{tho03}. In addition, the RASS luminosities
had 50 per cent uncertainties in addition to statistical errors after
conversion from the 0.1--2.4~keV to the 2--10~keV range.

The primary aim of this paper is to derive the most accurate shape of
the XLF to date by using a carefully selected sample of DNe. In order
to achieve this, we aim to minimise the biases seen in other published
XLFs by using sample selection criteria described in
Section~\ref{sourcesample}. One of the criteria worth mentioning here
is that we only use parallax-based distance estimates. The source
sample used in this study does not represent a complete sample of DNe
within 200~pc: the sample is more of a ''fair sample'' which was not
chosen based on the X-ray properties of the sources, and which
represents typical DNe within the solar neighbourhood. We will
consider the possible effects of the optical selection, inherent in
the parallax sample, on the XLF which we derive in Section 6
(Discussion). The motivation for choosing a group of DNe as the sample
is based on observations of the Galactic CV populations and previous
CV population models. Various authors, such as \citet{how97},
\citet{pre07a} and \citet{gan09}, have pointed out that binaries which
are brighter in X-rays and show frequent outbursts, may not represent
the true majority of Galactic CV population. The discovery methods of
CVs usually favour brighter systems, and thus fainter CVs with low
mass accretion rates are likely to be under-represented. However,
population models of CVs predict that the majority of CVs are
short-period systems ($P_{orb}$ $<$ 2.5 h) and X-ray faint
\citep[e.g.][]{kol93,how97}. These models are supported by
observational evidence, e.g. \citet{pat84} showed that a sample of CVs
with a total space density of 6 $\times$ 10$^{-6}$ pc$^{-3}$ was
dominated by low mass accretion rate, and thus short period,
systems. Also, the SDSS study by \citet{gan09} showed that orbital
periods of intrinsically faint Galactic CVs accumulated in the 80--86
min range; they found that 20 out of 30 SDSS CVs in this period range
showed characteristics which implied that they are low mass accretion
rate WZ Sge type DNe. As has been shown by these studies, less X-ray
luminous objects (such as DNe) dominated the studied volumes, and thus
we choose to focus on DNe in this paper. It is also worth mentioning
the study by \citet{pre07b} who carried out the {\it ROSAT} North
Ecliptic Pole (NEP) survey using a purely X-ray flux limited and a
complete sample of 442 X-ray sources above a flux limit of $\sim$
10$^{-14}$ erg cm$^{-2}$ s$^{-1}$ in the 0.5--2.0~keV band (only five
systems were CVs). They concluded that if the overall space density of
CVs is as high as 2 $\times$ 10$^{-4}$ pc$^{-3}$, then the dominant CV
population must be fainter than 2 $\times$ 10$^{29}$ erg s$^{-1}$.

We have carried out X-ray spectral analysis of our sample of 13
sources and derived an XLF in the 2--10 keV band for 12 of them with
reliable distance measurements based on those of by \citet{har04},
\citet{tho03} and \citet{tho08}. By using sources with accurate
distance measurements, we minimise the error on the luminosity. Also,
we have carried out timing analysis for 5 sources in the sample which
were recently observed by {\it Suzaku}. At the time of writing this
paper, the Z Cam type star KT Per went into an outburst in January
2009, and thus we also briefly report on the {\it Suzaku} observations
of KT Per during the outburst in Section~\ref{sec:ktper}.

\section{The selection criteria and the source sample}\label{sourcesample}
Since we wanted to obtain accurate luminosities for the sources (and
thus an accurate shape for the luminosity function), the first step
was to avoid selecting sources randomly from the archive \citep[see
e.g.][]{bas05} or selecting an X-ray flux limited source sample. The
aim was to have a {\it distance-limited} sample. Thus, sources were
not selected based on their X-ray properties, but we chose only those
DNe which have accurately measured distances based on trigonometric
parallax measurements within $\sim$ 200~pc. Note that by using all
available distance measurement techniques, Patterson (priv. comm.)
estimates that currently there are 13 DNe within 100~pc, and $\sim$ 33
DNe within 200~pc from the Sun, of which the latter count is clearly
incomplete. Above the 200~pc limit, ground-based parallax technique
does not give accurate and reliable distance measurements. By using
trigonometric parallax-based distance measurements, we are more likely
to avoid biases in the distance measurements which are present in the
previous, published X-ray luminosity functions. Due to the lack of
ground-based parallax measurement programme for the Southern
hemisphere, our sample is limited to northern and equatorial
objects. However, this selection should not introduce any biases in
terms of the optical or X-ray luminosities in our sample.

The distance measurements of the sources chosen for this work are
based on astrometric parallaxes obtained by the {\it Hubble Space
Telescope} (HST) Fine Guidance Sensors (FGSs) \citep{har04}, and
trigonometric parallaxes obtained by the ground-based 2.4 m Hiltner
Telescope at the MDM Observatory on Kitt Peak, Arizona
\citep[][]{tho03,tho08} and Thorstensen (in prep.). The first accurate
astrometric parallaxes of DNe (SS~Cyg, SS~Aur and U~Gem) were measured
in 1999 using the FGSs which can deliver high-precision parallaxes
with sub-milliarcsecond uncertainties \citep{har99}. Trigonometric
parallaxes derived by ground-based observations have uncertainties
around 1 mas (= 10$^{-3}$ arcsec) or less \citep[][]{tho03,tho08}
which is almost as good as the uncertainty on the FGS parallax
measurements.

The second selection criterion was to restrict the sample to sources
which had been observed by X-ray imaging telescopes with CCDs in the
energy range 0.2--10~keV. Once we had obtained a list of targets with
parallax measurements, we then looked for archival data of pointed
imaging X-ray observations of these targets in the energy range $\sim$
0.2--10~keV. If the chosen targets did not have previous X-ray imaging
observations, we requested {\it Suzaku} X-ray observations. Finally,
we wanted to constrain the sample to those sources which were in their
quiescent states during the observations in order to avoid biases in
the luminosities, and thus {\it AAVSO}\footnote{www.aavso.org} light
curves of the selected sources were inspected to confirm that the
sources were in quiescence during the X-ray observations.

The final source sample consists of 9 SU~UMa (including 2 WZ~Sge
systems), 3 U~Gem and 1 Z~Cam type DNe. The main characteristic which
separates these classes of DNe is the outburst behaviour: U~Gem type
DNe outburst mainly in timescales of every few weeks to every few
months whereas SU~UMa stars show normal, U~Gem type DN outbursts and,
in addition, superoutbursts with superhumps (variations in the light
curves at a period of a few per cent longer than the orbital period)
in timescales of several months to years. The extreme cases, WZ~Sge
stars, only have superoutbursts with outburst timescales of decades
without normal DN outbursts. The defining characteristic for Z~Cam
stars is standstills, i.e., it is possible that after an outburst,
they do not return to the minimum magnitude (unlike U~Gem stars), but
remain between the minimum and maximum magnitudes for 10--40 days.

\begin{table*}
\caption{The source sample used to derive the X-ray luminosity function (excluding *) with their inclinations, orbital periods, white
dwarf masses, distances and DN types. The types given in the last column are U Gem (UG), SU UMa (SU), WZ Sge (WZ) and Z Cam (ZC). The references are: a) \citet{tho03}, b) \citet{tho08}, c) \citet{har04}, d) \citet{mas01}, e) \citet{urb06}, f) \citet{fri90}, g) \citet{rit03}, h) \citet{hor91}, i) Preliminary distance estimate from Thorstensen (in prep.), and j) \citet{tem06}.}
\begin{tabular}{cccccc}
\\
 \hline
\hline
Source  & Inclination & P$_{orb}$ & M$_{WD}$ & Distance & Type \\
        & deg & h & M$_{\odot}$ & pc & \\
\hline
SS Cyg   &40 $\pm$ 8 $^{c}$ & 6.603 $^{c}$ & 1.19 $^{f}$ & 165$^{+13}_{-11}$ $^{c}$ & UG\\
V893 Sco &71 $\pm$ 5 $^{a}$ &1.82 $^{a}$ & 0.89 $^{d}$& 155$^{+58}_{-34}$ $^{a}$ & SU \\
SW UMa   &45 $\pm$ 18 $^{g}$ & 1.36 $^{b}$ &0.80 $^{e}$ & 164$^{+22}_{-19}$ $^{b}$ & SU \\
VY Aqr   &63 $\pm$ 13 $^{a}$ & 1.51 $^{a}$ & 0.8/0.55 $^{e}$ & 97$^{+15}_{-12}$ $^{a}$ & SU \\
SS Aur   &40 $\pm$ 7 $^{c}$ & 4.39 $^{a}$ & 1.03 $^{e}$ & 167$^{+10}_{-9}$ $^{c}$ & UG \\
BZ UMa   &60--75 $^{e}$ & 1.63 $^{b}$ & 0.55 $^{e}$ & 228$^{+63}_{-43}$ $^{b}$ & SU \\
U Gem    &69 $\pm$ 2 $^{c}$ & 4.246 $^{c}$ & 1.03 $^{e}$ & 100 $\pm$ 4 $^{c}$ & UG \\
T Leo    &47 $\pm$ 19 $^{a}$ & 1.42 $^{a}$ & 0.35 $^{e}$ & 101$^{+13}_{-11}$ $^{a}$ & SU  \\
WZ Sge   &76 $\pm$ 6 $^{c}$ & 1.36 $^{a}$ & 0.90 $^{e}$ & 43.5 $\pm$ 0.3 $^{c}$ & SU/WZ \\
HT Cas   &81 $\pm$ 1 $^{h}$ & 1.77 $^{b}$ & 0.8 $^{e}$ & 131$^{+22}_{-17}$ $^{b}$ & SU  \\
GW Lib   &11 $\pm$ 10 $^{a}$ & 1.28 $^{a}$ & 0.8 $^{e}$ & 104$^{+30}_{-20}$ $^{a}$ & SU/WZ \\
Z Cam$^{*}$ &65 $\pm$ 10 $^{a}$& 6.98 $^{a}$ & 1.21 $^{e}$ & 163$^{+68}_{-38}$ $^{a}$ & ZC \\
ASAS J0025 &--& 1.37 $^{j}$ &--&$\sim$ 175$^{+120}_{-40}$ $^{i}$ & SU \\
\hline
\hline
\label{sources}
\end{tabular}
\end{table*}

The sources, which were included in the calculation of the X-ray
luminosity function and when testing different correlations discussed
later in this paper, were observed with {\it Suzaku} (BZ~UMa, SW~UMa,
VY~Aqr, SS~Cyg, SS~Aur, V893~Sco, and ASAS J002511+1217.2), {\it
XMM-Newton} (U~Gem, T~Leo, HT~Cas and GW~Lib) and with {\it ASCA}
(WZ~Sge). {\it Suzaku} observations of BZ~UMa, SW~UMa, VY~Aqr, SS~Aur,
V893~Sco and ASAS~J0025 were requested as these observations were not
in the archive. \citet{muk09} discuss the {\it Suzaku} observations of
V893~Sco in more detail. We also included Z Cam in the source sample
since it has a parallax measurement, but it appeared to be in a
transition state during the observations. Thus, we have only reported
the results of the spectral analysis for Z~Cam, but excluded it when
calculating the X-ray luminosity function, and when testing
correlations between different parameters. The system parameters for
all the 13 sources are given in Table~\ref{sources}.

\section{Observations and data reduction}
The details of the {\it Suzaku}, {\it XMM}, and {\it ASCA}
observations are given in Table \ref{observations}, and the data
reduction methods are described in the following sections.

\begin{table*}
\caption{The observation dates and the instruments used in the observations for each source. The exposure times for the {\it Suzaku} sources have been obtained from the cleaned event lists, and the numbers in brackets for the {\it XMM} sources show exposure times after filtering high background flares. The last column corresponds to the optical state of the source during the observations.}
\begin{tabular}{ccccccc}
\\
\hline
\hline
Source   & ObsID & Instrument & T$_{start}$ & T$_{stop}$ & T$_{exp}$ & State \\
         &       &            &             &            &    ks     & \\
\hline
SS Cyg   & 400006010 & XIS/Suzaku & 2005-11-02 & 2005-11-02 & 39 & Q  \\
V893 Sco & 401041010 & XIS/Suzaku & 2006-08-26 & 2006-08-27 & 18 & Q \\
SW UMa   & 402044010 & XIS/Suzaku & 2007-11-06 & 2007-11-06 & 17 & Q \\
VY Aqr   & 402043010 & XIS/Suzaku & 2007-11-10 & 2007-11-11 & 25 & Q \\
SS Aur   & 402045010 & XIS/Suzaku & 2008-03-04 & 2008-03-05 & 19 & Q \\
BZ UMa   & 402046010 & XIS/Suzaku & 2008-03-24 & 2008-03-25 & 30 & Q \\
ASAS     & 403039010 & XIS/Suzaku & 2009-01-10 & 2009-01-11 & 33 & Q \\
J0025    & & & & & & \\
KT Per   & 403041010 & XIS/Suzaku  & 2009-01-12 & 2009-01-13 & 29 & OB \\
U Gem    & 0110070401 & MOS1/XMM   & 2002-04-13 & 2002-04-13 & 23(22.4) & Q \\
         & 0110070401 & MOS2/XMM   & 2002-04-13 & 2002-04-13 & 23(22.4) & Q \\
T Leo    & 0111970701 & PN/XMM     & 2002-06-01 & 2002-06-01 & 13(13) & Q  \\
HT Cas   & 0111310101 & PN/XMM     & 2002-08-20 & 2002-08-20 & 50(6.9) & Q  \\
GW Lib   & 0303180101 & PN/XMM     & 2005-08-25 & 2005-08-26 & 22(6.7) & Q  \\
WZ Sge   & 34006000 & GIS,SIS/ASCA & 1996-05-15 & 1996-05-15 & 85 & Q  \\
Z Cam    & 35011000 & GIS,SIS/ASCA & 1997-04-12 & 1997-04-12 & 41 & T \\
\hline
\hline
\label{observations}
\end{tabular}
\end{table*}

\subsection{Suzaku data reduction}
{\it Suzaku} \citep{mit07}, originally {\it Astro-E2}, was launched in
2005 and is Japan's 5th X-ray astronomy mission. In this paper, we
will focus on the X-ray Imaging Spectrometer (XIS) data. The XIS
consists of four sensors: XIS0,1,2,3 of which three (XIS0,2,3) contain
front-illuminated (FI) CCDs, and XIS1 contains a back-illuminated (BI)
CCD. The XIS0,2,3 are less sensitive to soft X-rays than XIS1 due to
the thin Si and SiO$_{2}$ layers on the front side of the XIS0,2,3
CCDs. Since November 9, 2006, the XIS2 unit has not been available for
observations. The {\it Suzaku} background is low and hardly affected
by soft proton flares often seen in {\it XMM} observations.

The unfiltered event lists of SS Cyg and V893 Sco were reprocessed
with \textsc{xispi} and screened in \textsc{xselect} with
\textsc{xisrepro} since the pipeline version for these observations
was older than v.2.1.6.15 which does not include correction for the
time- and energy-dependent effects in energy scale calibration. For
all the other {\it Suzaku} observations, the observations had been
processed by more recent pipeline versions and thus reprocessing was
not necessary. Pile-up was not a problem for our data since the source
count rates were safely below the pile-up limit (12 ct s$^{-1}$) for
point sources observed in the 'Normal' mode using Full
Window\footnote{http://heasarc.gsfc.nasa.gov/docs/suzaku/analysis/abc/abc.html}.
The {\it Suzaku} data reduction described below was carried out in a
similar manner for all the {\it Suzaku} observations. The cleaned
event lists were read into \textsc{xselect} in which X-ray spectra
were extracted for each source. Light curves were extracted for SW
UMa, BZ UMa, SS Aur, ASAS J0025, and VY Aqr for timing analysis
studies. To include 99 per cent of the flux and to obtain the most
accurate flux calibration, the spectra and light curves were extracted
using a source radius of 260'' (250 pixels). The backgrounds were
taken as an annulus centred on the source excluding the inner 4'
source region. The outer radii of the background annuli were
determined according to how close the calibration sources were to the
target. The response matrix files (RMFs) and ancillary response files
(ARFs) were created and combined within \textsc{xisresp} v.1.10. The
XIS0,2,3 source and background spectra and the corresponding response
files were summed in \textsc{addascaspec} to create the total XIS0,2,3
source and background spectra, and the total XIS0,2,3 response
file. For the light curves, the background areas were scaled to match
the source areas, and the scaled background light curves were then
subtracted from the source light curves in \textsc{lcmath}.

\subsection{XMM data reduction}
{\it XMM-Newton} \citep{jan01} is the cornerstone mission of the
European Space Agency (ESA). It has been operating since 1999 with
three X-ray cameras (EPIC pn, MOS1 and MOS2), the Optical Monitor
(OM), and the Reflection Grating Spectrometer (RGS) onboard. The X-ray
cameras cover the energy range 0.2--12.0~keV.

The data were obtained from the {\it XMM-Newton} Science Archive (XSA)
and the {\it XMM-Newton} data were reduced and analysed in the
standard manner using the {\it XMM-Newton} Science Analysis System
\textsc{sas} version 8.0.0. Each observation was checked for high
background flares in the range 10--12 keV using single pixel events
(\textsc{pattern == 0}). The high background flares were cut above
0.35 ct s$^{-1}$ for the MOS data and above 0.40 ct s$^{-1}$ for the
pn data. The source and background extraction regions were taken from
circular extraction areas avoiding any contaminating background
sources. The radii of the source regions were calculated by using the
\textsc{sas} task \textsc{region} in order to derive source extraction
radii which include $\sim$ 90 per cent of the source flux for each
source. The background extraction region (r$_{bg}$ = 130 arcsec) was
taken from the same chip as the source extraction region, or from an
adjacent chip in case of a crowded source chip. When extracting the
X-ray spectra, only well-calibrated X-ray events were selected for all
the sources, i.e. for the pn spectra single and double pixel events
with \textsc{pattern} $\leq$ 4 were chosen, and in order to reject bad
pixels and events close to CCD gaps, \textsc{FLAG} == 0 was used. For
the MOS, \textsc{pattern $\leq$ 12} and \#XMMEA\_EM were applied.

For most of the observations, we analysed the pn observations
only. Since the total effective area of the two EPIC MOS cameras is
nearly equal to the effective area of the EPIC pn, the MOS spectra
would not add any significant information to the pn spectra. The only
exception was U~Gem pn observation which had been obtained in Small
Window mode. Thus, we used the MOS1 and 2 data which had been obtained
in Large and Small window modes, respectively, and selected the
backgrounds from the surrounding CCDs. To form the total MOS spectrum
for U~Gem, the U~Gem MOS1 and 2 spectra were summed in
\textsc{addascaspec}. All the observations were checked in case of
pile-up by using the {\it XMM} \textsc{sas} task
\textsc{epatplot}. Pile-up did not occur in any of the observations,
but the source PSFs of HT Cas and T Leo were contaminated by
out-of-time (OoT) events, introduced by these two sources. Therefore,
the background regions in the HT~Cas and T~Leo observations were taken
from the adjacent chip in order to avoid the OoT events. These OoT
events were removed from the source X-ray spectra according to the
'\textsc{sas}
threads'\footnote{http://xmm2.esac.esa.int/sas/8.0.0/documentation/threads/\\EPIC\_OoT.html},
i.e. the source spectra extracted from the OoT event lists were
subtracted from the source spectra extracted from the original event
list.

\subsection{ASCA data reduction}
The {\it Advanced Satellite for Cosmology and Astrophysics}
\citep[{\it ASCA},][]{tan94} was Japan's fourth cosmic X-ray astronomy
mission operating between February 1993 and July 2000 and was the
first X-ray observatory which carried CCD cameras. The main science
goal of {\it ASCA} was the X-ray spectroscopy of astrophysical
plasmas. It carried four X-ray telescopes with two types of detectors
located inside them: two CCD cameras, i.e. the Solid-state Imaging
Spectrometers (SIS0 and SIS1) with spectral resolution of 2 per cent
at 5.9 keV at launch, and two scintillation proportional counters,
i.e. the Gas Imaging Spectrometers (GIS2 and GIS3).

The {\it ASCA} data reduction was performed in the standard manner by
mostly using the standard screening values for the GIS and SIS
instruments as described in NASA {\it ASCA} online
manual\footnote{http://heasarc.gsfc.nasa.gov/docs/asca/abc/abc.html}.
For both instruments, intervals outside the South Atlantic Anomaly
(SAA) were chosen, also including intervals when the attitude control
was stable with the upper limit of the angular distance from the
target set to \textsc{ang\_dist} $<$ 0.02 degrees. For the SIS
instruments, the bright earth angle of \textsc{br\_earth} $>$ 10 was
applied excluding the data taken below the 10$^{\circ}$ angle. Times
of high background were excluded when the PIXL monitor count rate was
3$\sigma$ above the mean of the observation. Also, the background
monitor count rate of \textsc{rbm\_cont} $<$ 500 was applied (the
standard screening value is \textsc{rbm\_cont} $<$ 100). Events which
occurred before the first Day-Night transition and at least 32 seconds
after the Day-Night transition, and also before the passage of the
South Atlantic Anomaly (SAA) and at least 32 s after the SAA, were
selected (\textsc{T\_DY\_NT < 0 $\parallel$ T\_DY\_NT > 32} \&\&
\textsc{T\_SAA < 0 $\parallel$ T\_SAA > 32}).

The source extraction regions for the SIS and GIS were centred on the
source. For the GIS, a $\sim$ 6 arcmin circular source extraction
region was used for both Z~Cam and WZ~Sge, while the SIS source
extraction regions were smaller so that they could be safely fitted
within the chip. Thus, the source extraction radii for Z~Cam were 4.4
arcmin (SIS0) and $\sim$ 3.5 arcmin (SIS1) and for WZ~Sge $\sim$ 4
arcmin (SIS0) and $\sim$ 3 arcmin (SIS1). The background extraction
region for the GIS was taken centred on the detector excluding a
circular region of $\sim$ 8 arcmin centred on the source. The
background extraction radii for GIS2 and GIS3 were 15.7 and 15.5
arcmin (Z~Cam), and 15.4 and 13.8 arcmin (WZ~Sge) respectively. For
the SIS background, blank-sky background observations were used for
Z~Cam due to lack of space for a local background region on the
CCDs. For WZ~Sge, the total area of the two active CCDs excluding a
5.5 arcmin region around the source was used as the background
extraction region.

The ancillary response files (ARFs) for the GIS and SIS spectra were
created with \textsc{ascaarf} and the SIS response matrix files (RMFs)
with \textsc{sisrmg}. Finally, the total SIS and GIS X-ray spectra
were created by combining SIS0 and SIS1, and the GIS2 and GIS3 spectra
in \textsc{addascaspec}, respectively.

\section{Timing analysis}
Since VY~Aqr, SS~Aur, BZ~UMa, SW~UMa and ASAS~J0025 have not been
subject to previous, pointed, imaging X-ray observations before the
{\it Suzaku} observations, we looked for periodicities from the data
of these sources. KT~Per has been observed by the {\it Einstein}
Observatory \citep{cor84}, but no previous X-ray spectral or timing
analysis studies have been carried out for it. In order to ensure that
these objects are not intermediate polars (IPs) and to look for
orbital and spin modulation in the data, the power spectra were
calculated by using a Lomb-Scargle periodogram \citep{sca82} which is
used for period analysis of unevenly spaced data. When searching over
the frequency range 0.00001--0.03 Hz, no significant periodicities
were seen at the 99 per cent confidence level.

\section{Spectral analysis} 
We carried out X-ray spectral analysis in order to study the
underlying spectra of the source sample, and, ultimately, to calculate
the fluxes and luminosities of the sources. To employ Gaussian
statistics, the X-ray spectra were binned at 20 ct bin$^{-1}$ with
\textsc{grppha} and then fitted in \textsc{Xspec11} \citep{arn96}.

In CVs, the power source of X-ray emission is known to be accretion
onto the white dwarf. The accreted material is shock-heated to high
temperatures \citep[$kT_{max}$ $\sim$ 10--50~keV,][]{muk03}, and this
material has to cool before settling onto the white dwarf
surface. Thus, the cooling gas flow is assumed to consist of a range
of temperatures which vary from the hot shock temperature $kT_{max}$
to the temperature of the optically thin cooling material which
eventually settles onto the surface of the white dwarf
\citep{muk97}. Thus, when fitting X-ray spectra of CVs, cooling flow
spectral models should represent more physically correct picture of
the cooling plasma, unlike single temperature spectral models. Cooling
flow models have successfully been applied to CV spectra in previous
studies by e.g. \citet{whe96} and \citet{muk03}. In this view, the
multi-temperature characteristic is our motivation for emphasizing the
cooling flow model in the rest of this work. The differential emission
measure dEM/dT for an isobaric cooling flow can be described by
\citep{pan05}

\begin{equation}\label{eq:cool}
\frac{dEM}{dT} = \frac{5k\dot{M}n^{2}}{2\mu m_{p} \epsilon(T,n)},
\end{equation}

where $m_{p}$ is the mass of a proton, $\mu$ the mean molecular weight
($\sim$ 0.6), $\epsilon$(T,n) total emissivity per volume in units of
erg s$^{-1}$ cm$^{-3}$, $\dot{M}$ accretion rate, $n$ particle
density, and $k$ the Boltzmann constant. The source of the X-ray
emission above the white dwarf illuminates the surface of the white
dwarf and thus causes a reflection, which is seen as Fe K$\alpha$ iron
fluorescence line at 6.4~keV \citep{geo91}. According to George \&
Fabian, an infinite slab reflector subtending a total solid angle of
$\Omega$ = 2$\pi$ where the X-ray source is located right above the
slab, produces an equivalent width of up to $\sim$ 150~eV for the
6.4~keV Fe K$\alpha$ fluorescence line. The equivalent width of the
6.4~keV iron line depends on the total abundance of the reflector
\citep{don97}, the inclination angle between the surface of the
reflector and the observer's line of sight, and the photon index of
the spectrum of the X-ray emission source \citep{ish09}.

\begin{table}
\centering
\caption{The equivalent widths of the Fe 6.4 keV line derived by using the absorbed optically thin thermal plasma and cooling flow models.}
\begin{tabular}{ccc}
\\
\hline
\hline
Name & EW(mekal) & EW(mkcflow) \\ 
     &  eV    & eV \\ 
\hline
BZ UMa & 67$^{+42}_{-42}$ & $<$ 79  \\
\\
HT Cas & $<$ 81 & $<$ 91  \\
\\
SS Aur & 73$^{+37}_{-36}$ & 86$^{+52}_{-53}$  \\
\\
SW UMa & 201$^{+124}_{-124}$ & $<$ 141  \\
\\
U Gem  & 50$^{+25}_{-24}$ & 60$^{+33}_{-32}$  \\
\\
T Leo  & 71$^{+38}_{-38}$ & $<$ 73 \\
\\
V893 Sco & 46$^{+12}_{-11}$ & 45$^{+11}_{-12}$ \\
\\
VY Aqr & $<$ 156 & $<$ 157 \\
\\
WZ Sge & $<$ 140 & $<$ 76 \\
\\
SS Cyg & 75$^{+9}_{-4}$ & 73$^{+6}_{-7}$ \\
\\
Z Cam  & 120$^{+42}_{-42}$ & 164$^{+42}_{-43}$ \\
\\
ASAS J0025 & $<$ 220 & $<$ 200 \\
\hline
\hline
\label{ev}
\end{tabular}
\end{table}

Even though we believe that the cooling flow -type multi-temperature
model is the correct description of the physics of the cooling gas
flow in CVs, previous works have often used single temperature plasma
models. Thus, in order to compare the effects of two different
spectral models on the spectral fit parameters, we fitted the spectra
with 1) a single temperature optically thin thermal plasma model
\citep[\texttt{mekal,}][]{mew86, lie95} and 2) a cooling flow model
(\texttt{mkcflow}) which was originally developed to describe the
cooling flows in clusters of galaxies \citep{mus88}, adding
photoelectric absorption \citep[\texttt{wabs},][]{mor83} to both
models. In order to investigate the equivalent width of the 6.4~keV
iron emission line, a Gaussian line was added at 6.4~keV with a line
width fixed at $\sigma$ = 10~eV. The spectral fits did not necessarily
require the 6.4~keV line, e.g., for SS~Aur the $\chi^{2}_{\nu}$/$\nu$
= 0.96/629 when a Gaussian line at 6.4~keV was not included.

\begin{table*}
\caption{The fit results of the absorbed optically thin thermal plasma model with a 6.4~keV Gaussian line. The errors are 90 per cent confidence limits on one parameter of interest. $n_{H_1}$ and $n_{H_2}$ are the absorption columns of the photoelectric absorption (\texttt{wabs}) and partial covering (\texttt{pcfabs}) models, respectively. CFrac is the covering fraction of the partial covering model, kT the plasma temperature and Ab the abundance.}
\begin{tabular}{cccccccc}
\\
\hline
\hline
Name & $n_{H_{1}}$ & $n_{H_{2}}$ & CFrac & kT & Ab  & $\chi^{2}_{\nu}$/$\nu$ & P$_{null}$ \\ 
     & 10$^{20}$ cm$^{-2}$ & 10$^{20}$ cm$^{-2}$ & & keV & $Z_{\odot}$ & & \\
\hline
BZ UMa & $<$0.19 &--&--& 4.17$^{+0.16}_{-0.17}$&0.51$^{+0.07}_{-0.08}$ & 1.02/905 & 0.349 \\ 
\\
GW Lib &3.16$^{+8.58}_{-3.16}$&--&--&1.62$^{+1.92}_{-0.68}$&0.20$^{+1.23}_{-0.20}$& 1.02/8 & 0.414 \\
\\
HT Cas &--&15.36$^{+2.13}_{-1.08}$ & 0.95$^{+0.04}_{-0.04}$ & 6.43$^{+0.60}_{-0.64}$ & 0.71$^{+0.18}_{-0.17}$ & 1.07/259 & 0.207 \\
\\
SS Aur & $<$0.56 &--&--&6.35$^{+0.40}_{-0.40}$&1.0$^{+0.14}_{-0.15}$ & 1.06/628 & 0.127 \\ 
\\
SW UMa & $<$0.20 &--&--&2.77$^{+0.12}_{-0.13}$&0.20$^{+0.07}_{-0.05}$ & 1.23/470 & 5.36$\times$10$^{-4}$ \\ 
\\
U Gem  &0.89$^{+0.19}_{-0.20}$&--&--&0.78$^{+0.03}_{-0.01}$&1.05$^{+0.12}_{-0.09}$&1.34/401 & 5.80$\times$10$^{-6}$ \\ 
       &                      &  &  &6.85$^{+0.22}_{-0.23}$&                      &         & \\
\\
T Leo  &1.09$^{+0.21}_{-0.21}$&--&--&3.55$^{+0.10}_{-0.11}$&0.50$^{+0.06}_{-0.06}$ & 1.40/631 & 8.36$\times$10$^{-11}$  \\ 
\\
V893 Sco &--&80.89$^{+4.18}_{-3.87}$&0.86$^{+0.01}_{-0.01}$&7.99$^{+0.29}_{-0.27}$&0.76$^{+0.04}_{-0.05}$& 1.02/1936 & 0.245 \\ 
\\
VY Aqr & $<$1.64 &--&--&5.06$^{+0.43}_{-0.50}$&0.66$^{+0.18}_{-0.17}$ & 0.90/445 & 0.942 \\ 
\\
WZ Sge &8.97$^{+2.41}_{-1.92}$&--&--&4.88$^{+0.55}_{-0.54}$&0.33$^{+0.17}_{-0.19}$&0.84/409 & 0.993 \\ 
\\
SS Cyg &2.98$^{+0.14}_{-0.25}$&--&--&10.44$^{+0.16}_{-0.17}$&0.51$^{+0.02}_{-0.01}$&1.24/2881 & 2.74$\times$10$^{-17}$ \\
\\
Z Cam  &28.21$^{+2.63}_{-2.62}$ &292.41$^{+99.42}_{-68.86}$&0.35$^{+0.04}_{-0.05}$&8.68$^{+0.84}_{-0.79}$& 1.0 & 1.10/769 & 0.03 \\
\\
ASAS & $<$0.84 &--&--& 4.38$^{+0.61}_{-0.53}$&0.56$^{+0.29}_{-0.24}$ & 0.88/366 & 0.958 \\
J0025 & & & & & & & \\
\hline
\hline
\label{mekal}
\end{tabular}
\end{table*}

\begin{table*}
\caption{The fit results of the absorbed cooling flow model with a 6.4 keV Gaussian line. The errors are 90 per cent confidence limits on one parameter of interest. $n_{H_1}$ and $n_{H_2}$ are the absorption columns of the photoelectric absorption (\texttt{wabs}) and partial covering (\texttt{pcfabs}) models, respectively. CFrac is the covering fraction of the partial covering model, kT$_{max}$ the shock temperature and Ab the abundance.}
\begin{tabular}{cccccccc}
\\
\hline
\hline
Name & $n_{H_{1}}$ & $n_{H_{2}}$ & CFrac & kT$_{max}$ & Ab & $\chi^{2}_{\nu}$/$\nu$ & P$_{null}$ \\ 
     & 10$^{20}$ cm$^{-2}$ & 10$^{20}$ cm$^{-2}$ &  & keV  &  $Z_{\odot}$   &     &  \\
\hline
BZ UMa & $<$0.87 &--&--& 13.71$^{+1.38}_{-0.81}$ & 0.57$^{+0.13}_{-0.07}$ & 0.88/904 & 0.994 \\ 
\\
GW Lib &$<$ 3.76 &--&--&6.96$^{+8.79}_{-3.12}$&1.0& 0.60/8 & 0.782 \\
\\
HT Cas &--&16.74$^{+4.05}_{-2.49}$ & 0.92$^{+0.05}_{-0.04}$ & 23.09$^{+4.15}_{-5.33}$ & 0.78$^{+0.27}_{-0.22}$ & 0.99/258 & 0.525 \\
\\
SS Aur &3.30$^{+1.79}_{-1.51}$&--&--& 23.47$^{+4.01}_{-3.02}$ &1.0& 0.95/628 & 0.832 \\ 
\\
SW UMa & $<$0.67 &--&--& 8.33$^{+0.62}_{-0.99}$ & 0.41$^{+0.08}_{-0.10}$ & 0.87/469 & 0.978 \\ 
\\
U Gem  &0.76$^{+0.28}_{-0.21}$&--&--&25.82$^{+1.98}_{-1.43}$&1.04$^{+0.13}_{-0.11}$& 1.23/402 & 1.1$\times$10$^{-3}$ \\ 
\\
T Leo  &0.68$^{+0.24}_{-0.21}$&--&--&10.97$^{+0.85}_{-0.67}$ & 0.50$^{+0.07}_{-0.07}$& 1.08/629 & 7.24$\times$10$^{-2}$ \\ 
\\
V893 Sco &--&103.71$^{+3.98}_{-3.07}$&0.90$^{+0.01}_{-0.01}$&19.32$^{+1.29}_{-1.40}$&0.94$^{+0.05}_{-0.05}$& 0.94/1934 & 0.973 \\ 
\\
VY Aqr &1.10$^{+3.15}_{-1.10}$&--&--& 16.47$^{+2.68}_{-2.22}$ & 0.69$^{+0.25}_{-0.20}$ & 0.86/444 & 0.984 \\ 
\\
WZ Sge &11.58$^{+3.96}_{-3.06}$&--&--&13.31$^{+3.01}_{-3.16}$ & 0.23$^{+0.16}_{-0.13}$ & 0.83/408 & 0.996 \\ 
\\
SS Cyg &2.84$^{+0.11}_{-0.11}$&--&--&41.99$^{+1.20}_{-0.76}$&0.61$^{+0.03}_{-0.02}$ & 1.19/2883 & 4.80$\times$10$^{-12}$ \\ 
\\
Z Cam  &31.92$^{+4.77}_{-5.00}$ & 180.28$^{+53.37}_{-35.14}$&0.47$^{+0.07}_{-0.06}$&25.76$^{+5.16}_{-2.39}$ & 1.0 & 1.08/768 & 0.06 \\
\\
ASAS & $<2.67$ &--&--& 14.43$^{+4.36}_{-2.69}$ & 0.68$^{+0.44}_{-0.29}$ & 0.81/366 & 0.996  \\
J0025 & & & & & & & \\
\hline
\hline
\label{mkcflow}
\end{tabular}
\end{table*}

The {\it Suzaku} XIS1 and XIS0,2,3 spectra were fitted simultaneously
for each source as well as the {\it ASCA} GIS and SIS spectra of Z~Cam
and WZ~Sge with the models mentioned above. Some data sets required
additional components to improve the fits. Three of the sources,
HT~Cas, V893~Sco and Z~Cam, required partial covering absorption
model, \texttt{pcfabs}, to reduce residuals in the low energy end
(between $\sim$ 0.6--2~keV). To reduce residuals around 0.80~keV in
the SS~Cyg spectrum, we added a Gaussian line at 0.81~keV with a line
width of 0.24~keV letting the line energy and width both to vary
free. For U~Gem, single absorbed optically thin thermal plasma model
yielded a $\chi^{2}$/$\nu$ = 2.23/403. Since the fit was not
statistically satisfactory, we added a second optically thin thermal
plasma component to improve the fit and obtained $\chi^{2}$/$\nu$ =
1.34/401 which was good enough for our analysis.

In the spectral fitting, the parameters of the spectral models were
tied between different instrument spectra but let to vary free, apart
from the Gaussian line energy at 6.4~keV and the line width $\sigma$
which were fixed. In order to estimate the abundances, the abundance
parameter of the models was let to vary free for most data sets. For
those sources for which abundance was significantly higher than the
solar value, it was fixed at 1.0. An example of a source with a
super-solar abundance is Z~Cam for which the obtained abundance was
1.46$^{+0.34}_{-0.19}$ $Z_{0}$ with the partial covering +
photoelectric absorption combined with the cooling flow model when the
abundance was let to vary free.

\begin{figure*}
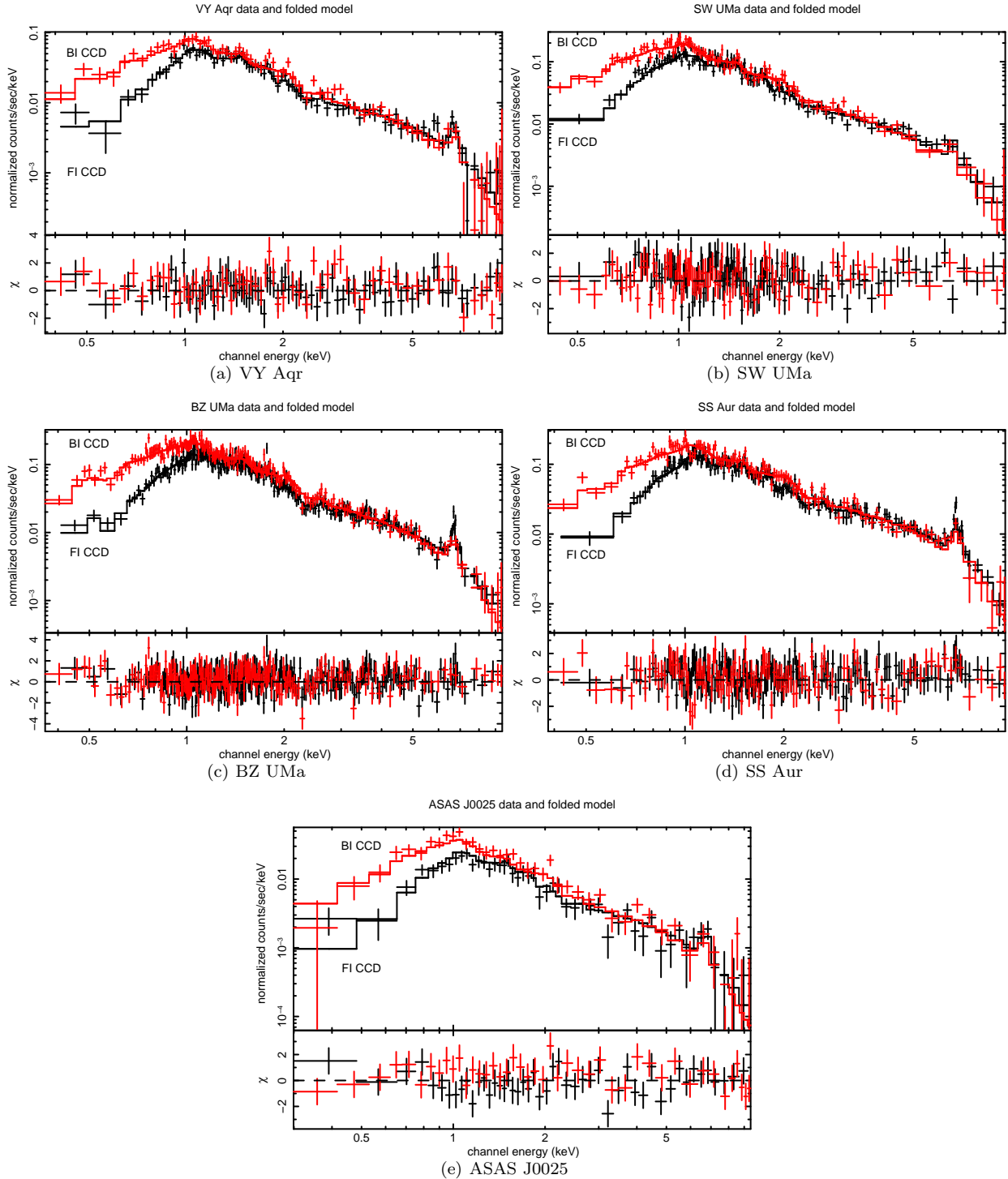

\subfigure[VY Aqr]{\label{vyaqr}\includegraphics[width=6cm,angle=-90]{fig1.eps}}
\subfigure[SW UMa]{\label{swuma}\includegraphics[width=6cm,angle=-90]{fig2.eps}}
\subfigure[BZ UMa]{\label{bzuma}\includegraphics[width=6cm,angle=-90]{fig3.eps}}
\subfigure[SS Aur]{\label{ssaur}\includegraphics[width=6cm,angle=-90]{fig4.eps}}
\subfigure[ASAS J0025]{\label{asas}\includegraphics[width=6cm,angle=-90]{fig5.eps}}
\linespread{1}
\caption{The X-ray spectra of (a) VY Aqr, (b) SW UMa, (c) BZ UMa, (d) SS Aur, and (e) ASAS J0025 fitted with an absorbed cooling flow model and a 6.4~keV Gaussian line (upper panels). The lower panel in each figure shows the residuals. The black spectra correspond to the front-illuminated (FI) XIS0,3 and the red ones to the back-illuminated (BI) XIS1 spectra.}
\label{spectra}
\end{figure*}

The measured equivalent widths of the 6.4~keV line for each source are
given in Table~\ref{ev}, and the results of the spectral fitting for
the optically thin thermal plasma and the cooling flow models are
given in Table~\ref{mekal} and \ref{mkcflow}, respectively. These
results show that in general, better $\chi^{2}_{\nu}$/$\nu$ values are
achieved with the cooling flow model. For example, the improvement
with the cooling flow model was statistically significant for SW~UMa
and T~Leo. Fig.~\ref{spectra} illustrates the X-ray spectra of the new
{\it Suzaku} XIS observations, i.e. VY~Aqr, SW~UMa, BZ~UMa, SS~Aur,
and ASAS J0025, which have been fitted with the cooling flow model
absorbed by photoelectric absorption with an added 6.4~keV Gaussian
line component. Most of the X-ray spectra show that the clearest,
discrete emission feature seen in the spectra of our source sample is
the iron Fe K$\alpha$ complex, except in GW~Lib, for which the
signal-to-noise at $\sim$ 6~keV is too low for a reliable measurement.

\subsection{Absorption}
Since the studied sources are all within $\sim$ 200~pc, i.e., within
the solar neighbourhood, the effect of interstellar absorption should
be negligible. Thus, high measured absorption columns would mainly be
due to intrinsic absorption, associated with the sources. For most of
the sources, the measured absorption columns were typically of the
order of a few $\times$ 10$^{20}$ cm$^{-2}$, or even lower (10$^{19}$
cm$^{-2}$) which indicate low intrinsic absorption.

The highest intrinsic absorption columns are found in V893~Sco, Z~Cam
and HT~Cas when compared to the rest of the source sample. All these
three sources have partial covering absorbers $n_{H_{2}}$ with values
of the order of 10$^{21}$ -- 10$^{22}$ cm$^{-2}$ depending on the
model. In addition to the partial covering absorber, Z Cam also has a
simple absorption component with the highest $n_{H_{1}}$ value,
$n_{H_{1}}$ $\sim$ 3 $\times$ 10$^{21}$ cm$^{-2}$, within the source
sample. Originally, V893~Sco was found to have high intrinsic
absorption by \citet{muk09}, and has a partial X-ray eclipse, also
discovered by their study. Also, according to the best-fit model of
\citet{bas01}, Z~Cam had large amounts of absorption with $n_{H}$ = 9
$\times$ 10$^{21}$ cm$^{-2}$ during the transition state. Baskill et
al. suggested that this absorption was associated with a clumpy disc
wind.

\subsection{Temperatures}
\begin{figure}
\centering
\includegraphics[width=80mm,angle=0]{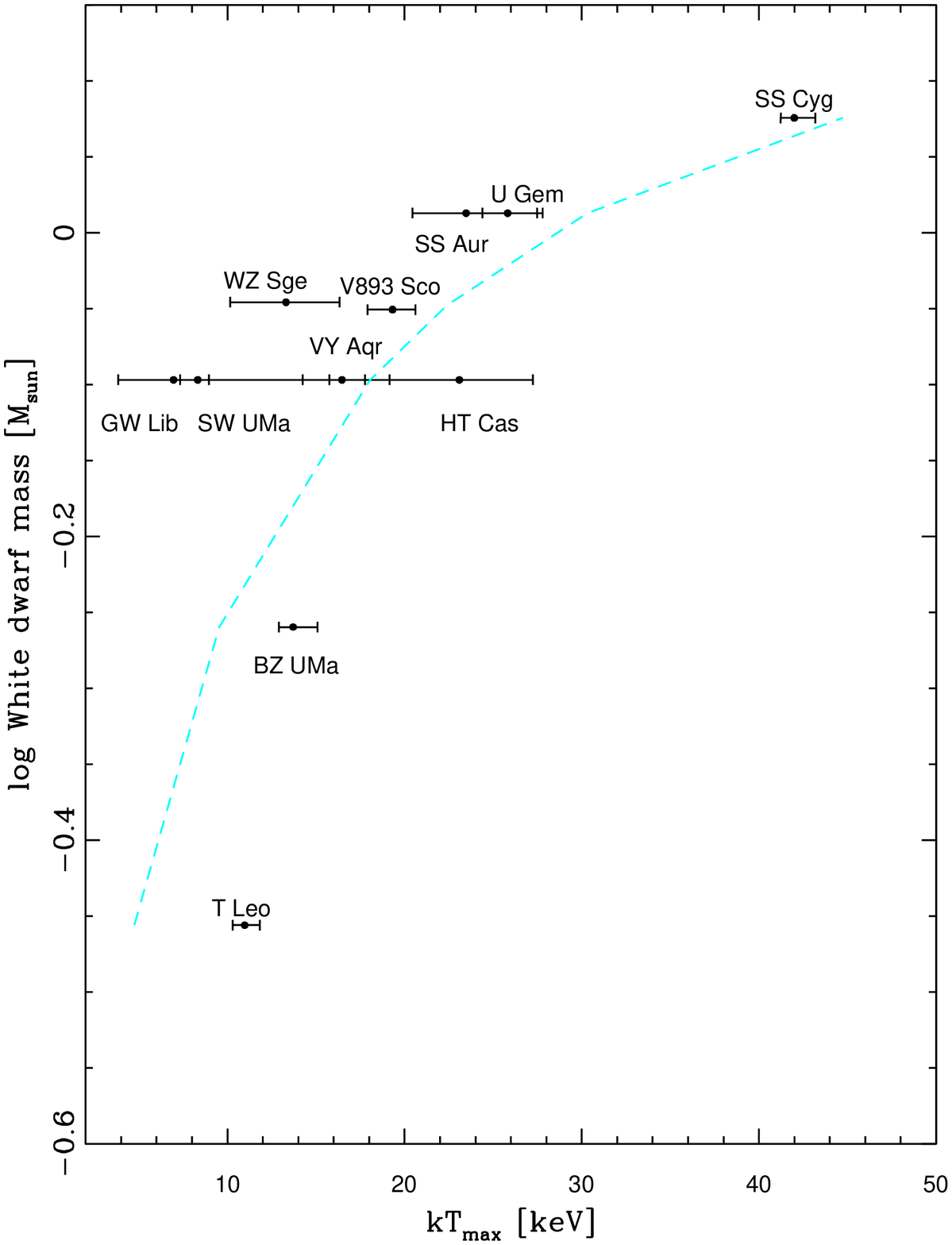}
\caption{Mass of the white dwarf versus the shock temperature,
$kT_{max}$, of the source sample with 90 per cent uncertainties for
$kT_{max}$. The light blue dashed line represents the theoretical
shock temperatures for given white dwarf masses. The figure does not
include ASAS J0025 since a mass estimate does not currently exist for
this source.}
\label{mass_temp}
\end{figure}

\begin{figure}
\centering
\includegraphics[width=70mm,angle=-90]{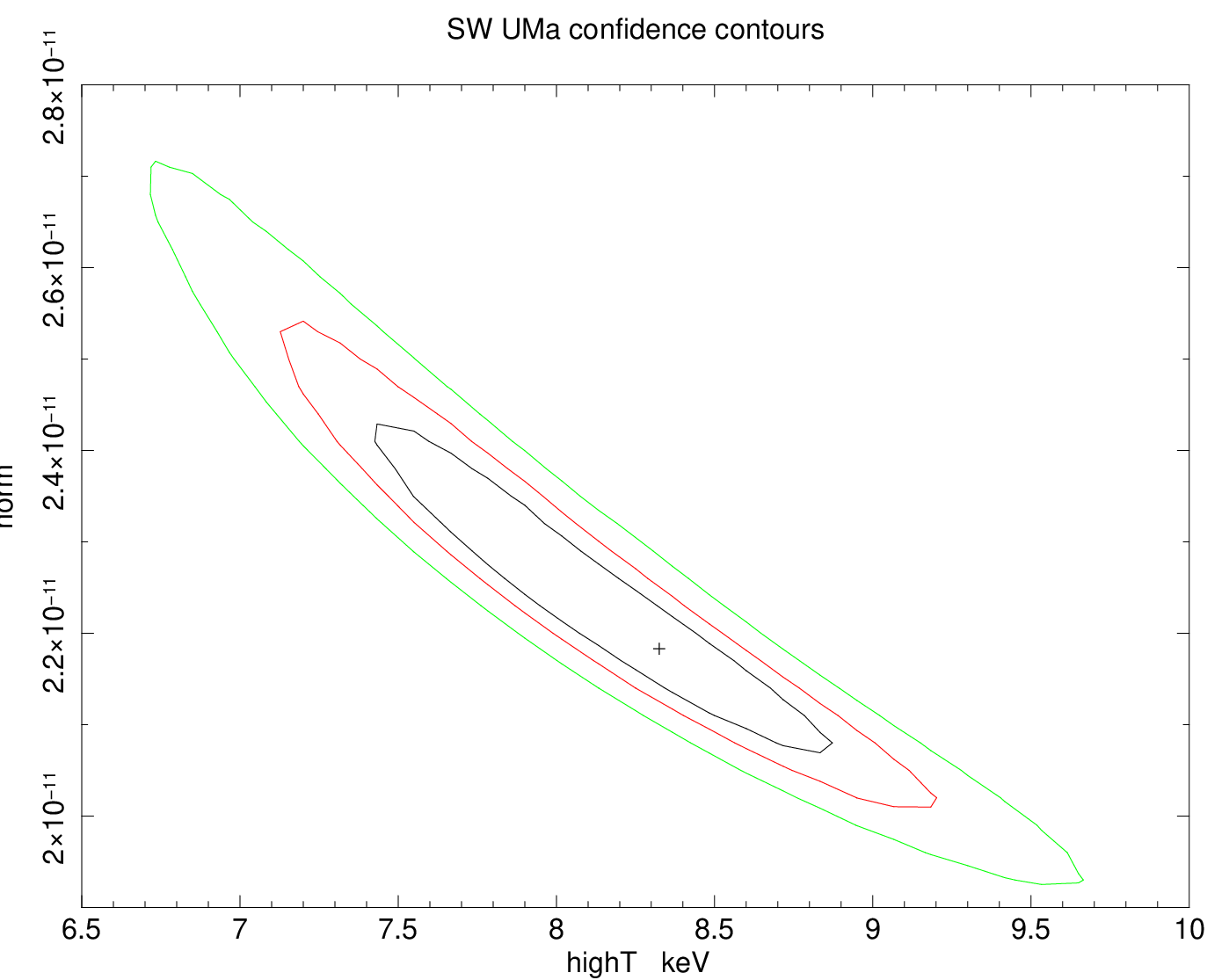}
\caption{The 68, 90, and 99 per cent confidence contours of SW~UMa for
the normalisation ($\dot{M}$) versus the shock temperature $kT_{max}$
of the cooling flow model.}
\label{2dim}
\end{figure}

The measured shock temperatures $kT_{max}$ seem to be correlated with
the white dwarf masses (Fig.~\ref{mass_temp}) as one would expect. In
Fig.~\ref{mass_temp} it has been assumed that the white dwarf mass of
VY~Aqr is 0.8 M$_{\odot}$ (see Table~\ref{sources}). ASAS J0025 is not
included in Fig.~\ref{mass_temp} since the mass estimate is currently
unknown. SS~Cyg appears to be located in the upper right corner due to
its high-mass white dwarf and thus high shock temperature. The shock
temperatures, $kT_{max}$, in Fig.~\ref{mass_temp} have been derived
from the spectral fits of the cooling flow model for each source. The
blue dashed line in Fig.~\ref{mass_temp} represents the theoretical
shock temperatures for given white dwarf masses. The radii, $R_{*}$,
of the given white dwarf masses, $M_{1}$, were calculated assuming the
mass-radius relation for cold, non-rotating and non-relativistic
helium white dwarfs \citep[see][]{pri75}

\begin{equation}
R_{*} = 7.7 \times 10^{8} x^{0.3767 - 0.00605\, \log(x)}\quad(cm),
\end{equation}

where $x$ = $\frac{1.44 M_{\odot}}{M_{1}}$ -1. Subsequently, the
theoretical shock temperatures, $T_{shock}$, for non-magnetic CVs were
calculated according to Eq.~\ref{eq:tshock} for optically thin gas
\citep{fra02}

\begin{equation}\label{eq:tshock}
T_{shock} = \frac{3}{16}\frac{GM_{1}\mu m_{H}}{kR_{*}},
\end{equation}

where $m_{H}$ is the mass of a hydrogen atom, $\mu$ the mean molecular
weight, and $k$ the Boltzmann constant. As it appears from
Fig.~\ref{mass_temp}, sources with high shock temperatures and low
luminosities are not seen. This is sensible since the X-ray luminosity
is proportional to $kT_{max}$ and the mass accretion rate, i.e. the
normalization of the cooling flow model (Eq.~\ref{eq:cool}), thus we
would expect to see high shock temperatures and high
luminosities. Also, due to this proportionality, we expect to see an
anti-correlation between $\dot{M}$ and $kT_{max}$ which indeed is seen
for example in SW~UMa (Fig.~\ref{2dim}).

As it appears from Fig.~\ref{mass_temp}, the white dwarf mass obtained
for T~Leo by \citet{urb06} is only 0.35 M$_{\odot}$. This mass
estimate may be unreliable, since as \citet{lem93} argue, a low white
dwarf mass would not allow superhumps to develop. See also
\citet{pat05} who refer to previous superhump studies which have shown
that the mass ratio $q$ = $M_{2}/M_{1}$ has a key role in producing
superhumps where $q_{crit}$ $\sim$ 0.3, although this value has not
been confirmed by observational evidence.

\subsection{Abundances}
We found that for most of the objects in the sample the obtained
abundances were sub-solar with both models. In general, the abundances
seem to be dependent on the spectral model: abundances are slightly
lower when the spectra are fitted with the optically thin thermal
plasma model. This is due to the single temperature characteristic of
the optically thin thermal plasma model, i.e. it is likely that the
abundances are underestimated because the best-fit temperature usually
converges close to the peak of the 6.7~keV He-like Fe K$\alpha$
emissivity, whereas the cooling flow model consists of a range of
temperatures outside the peak \citep[see][]{muk09}.

\subsection{X-ray fluxes and luminosities}
\begin{figure}
\centering
\includegraphics[width=80mm,angle=0]{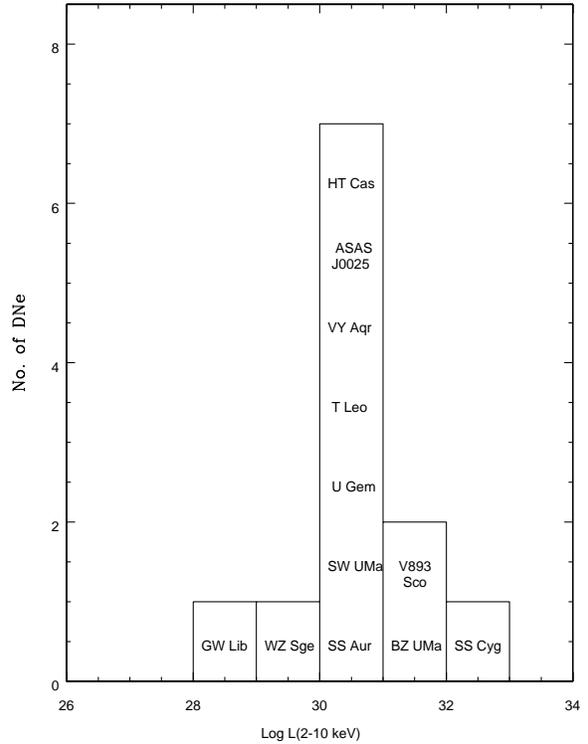}
\caption{A histogram showing the X-ray luminosities of the source sample in 2--10~keV.}
\label{histo}
\end{figure}

The 2--10 and bolometric 0.01--100~keV fluxes and luminosities which
were derived using the cooling flow model are given in Table
\ref{fluxes}. This shows that most of the 2--10~keV X-ray luminosities
are concentrated around 10$^{30}$ erg s$^{-1}$. This is also seen in
Fig.~\ref{histo} which shows a histogram of the X-ray luminosities of
our sample. Only one object, GW~Lib, stands out with a very low
luminosity (4 $\times$ 10$^{28}$ erg s$^{-1}$). The measured
luminosity of GW~Lib is consistent with the results obtained by
\citet{hil07}. \citet{byc09} showed that GW~Lib was still an order of
a magnitude brighter (L $\sim$ 10$^{30}$ erg s$^{-1}$) in X-rays
during {\it Swift} observations two years after the 2007 outburst than
in 2005. But since the optical magnitude had not reached the
quiescence level (V $\sim$ 18) in 2009, we do not consider the {\it
Swift} 2009 X-ray luminosity as the quiescent luminosity. Thus, the
higher X-ray luminosity measured in the {\it Swift} data does not
affect the results of this paper.

\begin{table*}
\caption{Fluxes and luminosities in the 2--10 (absorbed) and 0.01--100~keV (unabsorbed) bands derived from the cooling flow model for each source.}
\begin{tabular}{ccccc}
\\
\hline
\hline
Source & F(2--10 keV) & L(2--10 keV) & F(0.01--100 keV) & L(0.01--100 keV) \\  
       & x 10$^{-12}$ erg cm$^{-2}$ s$^{-1}$ & $\times$ 10$^{30}$ erg s$^{-1}$ & x 10$^{-12}$ erg cm$^{-2}$ s$^{-1}$ & $\times$ 10$^{30}$ erg s$^{-1}$ \\
\hline
BZ UMa     &2.4$^{+0.1}_{-0.2}$  & 14.9$^{+10.8}_{-5.9}$   & 5.8   & 36.5 \\
\\
GW Lib     &0.04$^{+0.03}_{-0.01}$  & 0.05$^{+0.10}_{-0.02}$  & 0.1  & 0.1 \\
\\
HT Cas     &2.9$^{+0.7}_{-0.4}$  & 6.1$^{+4.2}_{-2.2}$  & 7.2   & 14.9 \\
\\
SS Aur     &2.9$^{+0.3}_{-0.3}$  & 9.6$^{+2.3}_{-1.9}$  & 7.1   & 23.9 \\
\\
SW UMa     &1.5$^{+0.2}_{-0.1}$  & 4.9$^{+2.2}_{-1.3}$  & 4.2   & 13.7 \\
\\
U Gem      &6.9$^{+0.3}_{-0.3}$  & 8.3$^{+1.0}_{-1.0}$  & 17.1  & 20.6 \\
\\
T Leo      &5.2$^{+0.3}_{-0.3}$  & 6.4$^{+2.3}_{-1.7}$  & 13.2  & 16.3 \\
\\
V893 Sco   &17.3$^{+1.1}_{-1.1}$ & 50.1$^{+51.9}_{-21.4}$  & 45.7  & 133.0\\
\\
VY Aqr     &1.1$^{+0.2}_{-0.2}$  & 1.3$^{+0.8}_{-0.5}$  & 2.6  & 3.0 \\
\\
WZ Sge     &3.1$^{+1.2}_{-0.5}$  & 0.7$^{+0.3}_{-0.1}$  & 7.7  & 1.8 \\
\\
SS Cyg     &45.7$^{+0.5}_{-0.8}$ & 150.0$^{+29.0}_{-20.0}$ & 131.7 & 433.0 \\
\\
Z Cam      &19.2$^{+1.9}_{-2.6}$ & 61.6$^{+74.4}_{-30.3}$  & 52.3  & 168.0 \\
\\
ASAS J0025 &0.4$^{+0.1}_{-0.1}$  & 1.6$^{+3.8}_{-0.8}$  & 1.1  & 3.9 \\
\hline
\hline
\label{fluxes}
\end{tabular}
\end{table*}

One of the sources in our sample, SS~Aur, has previously been listed in
the {\it RXTE} All-Sky Slew Survey catalog where it appears more
luminous in X-rays than in our {\it Suzaku} observation (the {\it
RXTE} flux of SS Aur in 2--10~keV is $\sim$ 1.1 $\times$ 10$^{-11}$
erg cm$^{-2}$ s$^{-1}$). We suspect that the higher flux in the {\it
RXTE} observation is due to other, bright sources in the field which
overestimate the flux. E.g., the {\it ROSAT} Bright Source Catalogue
lists a cluster of galaxies, Abell 553, which is 53' away from SS Aur
and has a
WebPIMMS\footnote{http://heasarc.gsfc.nasa.gov/Tools/w3pimms.html}
estimated flux of $\sim$ 9.3 $\times$ 10$^{-12}$ erg cm$^{-2}$
s$^{-1}$ in 2--10~keV \citep[bremsstrahlung kT = 4~keV, Galactic
n$_{H}$ = 1.56 $\times$ 10$^{21}$ cm$^{-2}$ as in][]{ebe96}. Thus, the
higher {\it RXTE} flux of SS~Aur is very likely biased by the
background sources and not reliable.

\subsection{The outburst of KT Per}\label{sec:ktper}
\begin{figure*}
\centering
\includegraphics[width=70mm,angle=-90]{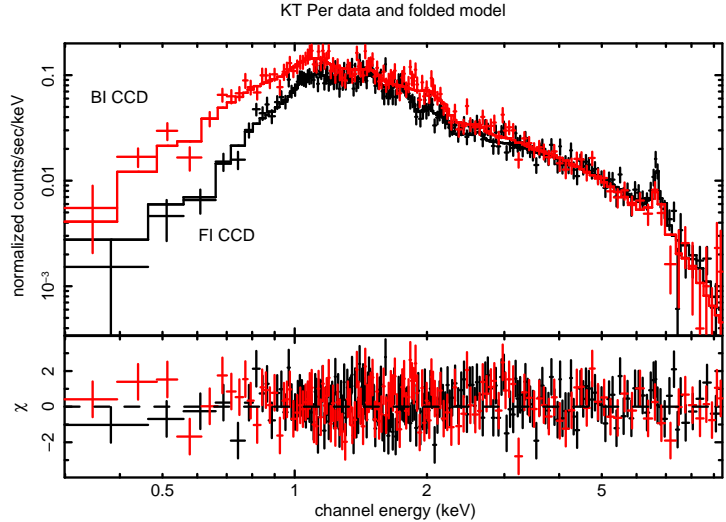}
\caption{The X-ray spectrum of KT~Per fitted with an absorbed cooling flow model and a 6.4~keV Gaussian line. The lower panel shows the residuals. The black spectrum corresponds to the front-illuminated (FI) XIS0,3 and the red one to the back-illuminated (BI) XIS1 spectrum.}
\label{ktper_fig}
\end{figure*}

We also analysed the {\it Suzaku} outburst data of KT~Per obtained in
January 2009, and report the results here. KT~Per is a Z~Cam type
dwarf nova, and was seen as an X-ray source by the {\it Einstein}
satellite in 1979 \citep{cor84}. We employed the same models which
were used for the source sample above, i.e. an absorbed optically thin
thermal plasma model and an absorbed cooling flow model with an added
6.4~keV line. Both models yielded acceptable fits:
$\chi_{\nu}^{2}/\nu$ = 0.97/838 (thermal plasma) and
$\chi_{\nu}^{2}/\nu$ = 0.95/837 (cooling flow). Fig.~\ref{ktper_fig}
shows the XIS1 and the combined XIS0,3 X-ray spectra of KT~Per which
have been fitted with an absorbed cooling flow model with a 6.4~keV
Gaussian line. The spectral fit parameters for the optically thin
thermal plasma and the cooling flow models with fluxes, luminosities
and fit statistics are given in Table~\ref{ktper}. The luminosities
given in Table~\ref{ktper} are calculated for the distance of
180$^{+36}_{-28}$ pc \citep{tho08}. \citet{bas05} noted that cooling
flow models are often a good representation of quiescent X-ray spectra
of CVs \citep[see also][]{muk03}, but not outburst spectra. Baskill et
al. applied the \textsc{Xspec} multi-temperature model \texttt{cevmkl}
to their {\it ASCA} spectra in order to fit a range of outburst and
quiescent spectra with a single simple model. We also investigated how
this multi-temperature model combined with photoelectric absorption
and a 6.4~keV Gaussian line would fit the outburst data of KT~Per, and
obtained a statistically acceptable fit: $\chi_{\nu}^{2}/\nu$ =
0.96/836, P$_{0}$ = 0.819.

\begin{table}
\centering
\caption{The fit parameters of KT~Per derived by using an absorbed
optically thin thermal plasma and absorbed cooling flow models with a
6.4~keV iron line. The errors are 90 per cent errors for one parameter
of interest.}
\begin{tabular}{ccc}
\\
\hline
\hline
Parameter & Thermal plasma & Cooling flow \\
\hline
n$_{H}$ & 14.40$^{+1.60}_{-1.54}$ & 15.70$^{+2.05}_{-1.78}$ \\
$\times$ 10$^{20}$ cm$^{-2}$ &  & \\
\\
kT  & 5.11$^{+0.32}_{-0.31}$ & 12.60$^{+1.47}_{-2.29}$ \\
(keV)\\
\\
Abundance & 0.40$^{+0.07}_{-0.07}$ & 0.42$^{+0.09}_{-0.08}$ \\
\\
EW  & 52$^{+39}_{-38}$ & 45$^{+44}_{-44}$ \\
(eV) & & \\
\\
Flux(2--10 keV) & 2.55 & 2.62   \\
$\times$ 10$^{-12}$ erg cm$^{-2}$ s$^{-1}$ & & \\
\\
Flux(0.01--100 keV) & 5.60 & 6.19 \\
$\times$ 10$^{-12}$ erg cm$^{-2}$ s$^{-1}$ & & \\
\\
Luminosity(2--10 keV) & 1.0 & 1.03 \\
$\times$ 10$^{31}$ erg s$^{-1}$ \\
\\
Luminosity(0.01--100 keV)& 2.19 & 2.42 \\
$\times$ 10$^{31}$ erg s$^{-1}$ & & \\
\\
$\chi^{2}_{\nu}/\nu$ & 0.97/838&0.95/837 \\
\\
P$_{0}$ &0.730&0.819 \\
\hline
\hline
\label{ktper}
\end{tabular}
\end{table}

\subsection{Calculating the height of the X-ray emission source above the white dwarf surface}
The method which is used for calculating the height of the X-ray
emission source above the white dwarf, has been explained by
\citet{ish09} for SS~Cyg. We have adopted the same method here in our
work. As was explained in the beginning of Section~5, an equivalent
width of up to $\sim$ 150~eV can be expected for the fluorescent Fe
K$\alpha$ line at 6.4~keV. In this work, we have assumed that the
reflection originates from the white dwarf surface only, thus the
reflection from the accretion disc is $\Omega_{disc}$/2$\pi$ = 0. The
equivalent width of 150~eV calculated by \citet{geo91} was assumed
under the solar abundance conditions of \citet{mor83} where [Fe/H] =
3.2 $\times$ 10$^{-5}$. We have employed the abundances of
\citet{and89} which are the default abundance values built in the
\textsc{Xspec} cooling flow and optically thin thermal plasma emission
models. For the solar abundances of Anders \& Grevesse, the [Fe/H]
composition is 4.68 $\times$ 10$^{-5}$. \citet{ish09}, who also
employed the Anders \& Grevesse abundances, corrected this abundance
difference (see their Eq.3) using their measured iron abundance of
0.37 $Z_{0}$. For solar abundance, the observed equivalent width of
the 6.4~keV line is

\begin{align}\label{eq:ew}
EW_{observed} &= 150 \times \frac{4.68 \times 10^{-5}}{3.2 \times 10^{-5}} \Big(\frac{\Omega_{WD}}{2\pi}\Big)Z\quad(eV)
                \nonumber \\
              &= 220 \Big(\frac{\Omega_{WD}}{2\pi}\Big)Z\quad(eV),
\end{align}

where $Z$ is the measured elemental abundance in solar units
$Z_{\odot}$ and $\Omega_{WD}$ the solid angle of the white dwarf
viewed from the plasma of the boundary layer. In our sample, the
observed equivalent widths (Table~\ref{ev}) are mainly below 150~eV
and this implies that the X-ray source is located at a height $h$
above the white dwarf surface. In the following, we use the values of
$EW_{observed}$ and abundances calculated from the cooling flow
model. As an example, the $EW_{observed}$ for SS~Aur is 86~eV and the
abundance is 1.0, thus Eq.~\ref{eq:ew} gives $\Omega_{WD}$/2$\pi$ =
0.39. If the X-ray source is point-like, the height $h$ of the X-ray
source above the white dwarf of a radius $R_{WD}$ is $h$ $<$ 0.14
$R_{WD}$. As another example, we obtain $\Omega_{WD}$/2$\pi$ = 0.22
and $h$ $<$ 0.64 $R_{WD}$ for V893~Sco ($EW_{observed}$ = 45~eV, $Z$ =
0.94 Z$_{0}$).

\section{Discussion}
We have derived an X-ray luminosity function for 12 dwarf novae using
archival {\it Suzaku}, {\it XMM-Newton}, and {\it ASCA} observations,
and obtained new observations for BZ~UMa, SW~UMa, VY~Aqr, SS~Aur,
V893~Sco and ASAS~J0025 with {\it Suzaku} as originally, they were not
available in the archive. Our results show that the 2--10~keV
luminosities, presented in Table~\ref{fluxes}, span a range between 4
$\times$ 10$^{28}$ and 1.5 $\times$ 10$^{32}$ erg s$^{-1}$, and that
most of the source luminosities in the sample are located within
10$^{30}$ erg s$^{-1}$, see Fig.~\ref{histo}, whereas, the X-ray
luminosities of the {\it ASCA} sample by \citet{bas05} were mainly
concentrated on higher luminosities between 10$^{31}$ and 10$^{32}$
erg s$^{-1}$. This difference is most likely due to the fact that we
did not apply X-ray selection criteria to our sample. Also, the
objects observed by {\it ASCA} were known to be X-ray bright, thus the
sample of Baskill et al. is very likely biased by sources which are
X-ray bright.

\begin{figure}
\centering
\includegraphics[width=90mm,angle=0]{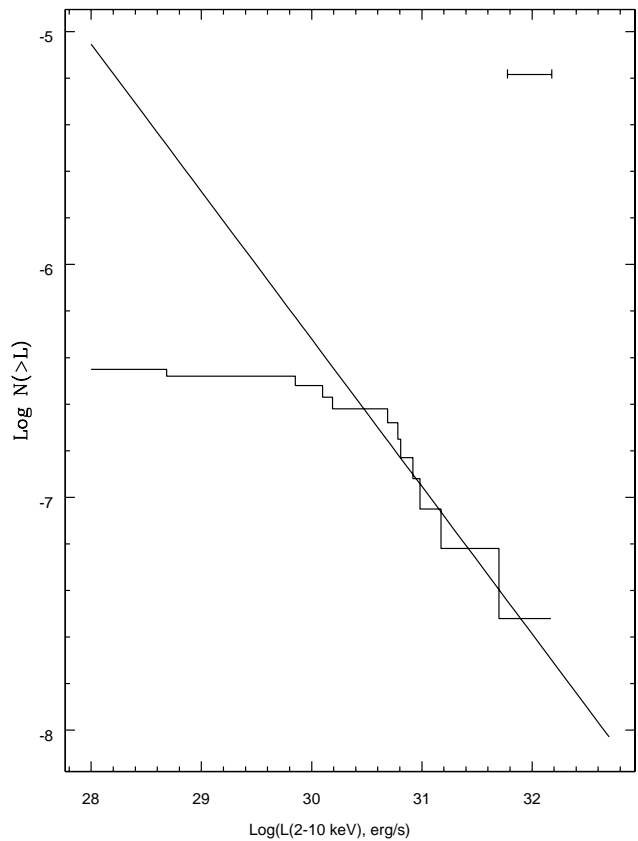}
\caption{The cumulative source distribution (histogram) and the
integrated power law luminosity function N($>$ L) as a function of
X-ray luminosity in log L in the 2--10~keV energy band. The error bar
on the top right represents a typical error on the luminosities.}
\label{lum}
\end{figure}

In order to derive the integrated X-ray luminosity function (XLF),
N($>$ L), for 12 sources within a distance of d = 200~pc, we assumed
that the luminosity function is characterized by a power law N($>$ L)
= k(L/L$_{t}$)$^{-\alpha}$ (see Fig.~\ref{lum} where the best-fit
parameters $\alpha$ = -0.64 and k = 2.39 $\times$ 10$^{-7}$,
corresponding to a threshold luminosity of L$_{t}$ = 3 $\times$
10$^{30}$ erg s$^{-1}$). The histogram illustrates the cumulative
source distribution per pc$^{3}$ in which a break is seen at $\sim$ 3
$\times$ 10$^{30}$ erg s$^{-1}$. This can be due to two possible
scenarios: 1) a single $\alpha$ power law describes the luminosity
function of DNe, but the sample becomes more incomplete below $\sim$ 3
$\times$ 10$^{30}$ erg s$^{-1}$ than it is above this limit, or 2) the
shape of the true XLF of DNe is a broken power law with a break at
around 3 $\times$ 10$^{30}$ erg s$^{-1}$. From these two scenarios,
the first one is more likely since the sample contains only a few
sources below $\sim$ 10$^{30}$ erg s$^{-1}$. Also, as was shown by,
e.g., the study of \citet{gan09}, more fainter CVs, such as WZ~Sge
types, are expected to exist. Based on the obtained power law slope,
the sample is dominated by the brighter DNe: this is probably caused
by the parallax measurement method which favours optically brighter
DNe which usually have high X-ray luminosities.

When calculating the total, integrated luminosity of the sample, we
restricted the calculations to the distance of 200~pc, thus excluding
BZ~UMa. Integrating between the luminosities of 1 $\times$ 10$^{28}$
and the maximum luminosity of the sample ($L_{max}$ = 1.50 $\times$
10$^{32}$ erg s$^{-1}$), yields the total integrated luminosity of
1.48 $\times$ 10$^{32}$ erg s$^{-1}$, whereas the integrated
luminosity between the threshold luminosity 3 $\times$ 10$^{30}$ and
$L_{max}$ is 1.15 $\times$ 10$^{32}$ erg s$^{-1}$. These two results
show that there are uncertainties in the integrated luminosities, most
likely caused by the small number of sources in the sample. In order
to obtain more accurate value for the integrated luminosity, the power
law slope ($\alpha$ = -0.64) should be better established. If the
obtained slope is not far from the true power law slope of DNe in the
solar neighbourhood, estimating the integrated luminosity more
accurately and constraining the bright luminosity end (10$^{32}$ erg
s$^{-1}$) requires more DNe to be included in the sample. Since the
source density at $\sim$ 10$^{32}$ erg s$^{-1}$ is $\sim$ 3 $\times$
10$^{-8}$ pc$^{-3}$ according to Fig.~\ref{lum}, we would need to
survey within a volume of 1 $\times$ 10$^{9}$ pc$^{3}$ to find $\sim$
30 SS~Cyg -type DNe and thus find a statistically significant
constraint for the brighter luminosities in the sample. This volume
would correspond to a distance of $\sim$ 620~pc with a flux limit of
$\sim$ 3.2 $\times$ 10$^{-12}$ erg cm$^{-2}$ s$^{-1}$.

Following this, we estimated how easy it would be to hide typical DN
luminosities in the solar neighbourhood. Assuming a typical dwarf nova
with a 5 keV bremsstrahlung and a low Galactic n$_{H}$ = 1 $\times$
10$^{20}$ cm$^{-2}$ in \textsc{WebPIMMs} yields a 2--10~keV flux of 5
$\times$ 10$^{-13}$ erg cm$^{-2}$ s$^{-1}$, corresponding to the {\it
ROSAT} PSPC count rate of 0.04 ct s$^{-1}$ which is just below the
detection limit (0.05 ct s$^{-1}$) of {\it ROSAT} PSPC
\citep{vog99}. Thus, luminosities above 2.4 $\times$ 10$^{30}$ erg
s$^{-1}$ within 200~pc or above 6 $\times$ 10$^{29}$ erg s$^{-1}$
within 100~pc should have been found by {\it ROSAT} and thus should be
in the RASS. However, given that sources with luminosities of
10$^{30}$ erg s$^{-1}$ and below at a distance of 100~pc were too
faint for the RASS, and that our XLF peaks at $\sim$ 10$^{30}$ erg
s$^{-1}$, we conclude that there is no existing X-ray selected sample
that we can use for this line of research.

How far is the total luminosity of our sample from accounting for the
total CV X-ray emissivity? In order to estimate this, we calculated
the absolute lower limit for the luminosity per cubic parsec volume
(L$_{x}$/vol). For a distance of r = 200~pc, the volume V = 4/3
$\times$ $\pi$ $\times$ (200 pc)$^{3}$ = 3.3 $\times$ 10$^{7}$
pc$^{3}$, and the total summed luminosity L$_{x}$ of the sample is
2.39 $\times$ 10$^{32}$ erg s$^{-1}$ (without BZ~UMa). Thus, the total
absolute lower limit L$_{x}$/volume = 7.24 $\times$ 10$^{24}$ erg
s$^{-1}$ pc$^{-3}$. Normalising this value to the local stellar mass
density 0.04 M$_{\odot}$ pc$^{-3}$ \citep{jah97} yields 1.81 $\times$
10$^{26}$ erg s$^{-1}$ M$_{\odot}^{-1}$ in the 2--10~keV range. For
comparison, \citet{saz06} obtained (1.1 $\pm$ 0.3) $\times$ 10$^{27}$
erg s$^{-1}$ M$_{\odot}^{-1}$ (2--10~keV) for the total CV X-ray
emissivity per unit stellar mass. Thus, our sample would account for
$\sim$ 16 per cent of this value.

And finally, how much would our sample account for the GRXE? The
Galactic Ridge X-ray emissivity estimated by \citet{rev06} in the
3--20~keV range was L$_{x}$/M $\sim$ (3.5 $\pm$ 0.5) $\times$
10$^{27}$ erg s$^{-1}$ M$_{\odot}^{-1}$, meaning that our sample would
account for 5 per cent of the Galactic Ridge X-ray emissivity. As we
estimated the X-ray emissivity of all CVs within 200~pc, we used the
exponential vertical density profile

\begin{equation}\label{density}
\rho(z) = \rho_{0}e^{|z|/h},
\end{equation}

of CVs with a scale height for short period systems ($h$ = 260~pc) as
in \citet{pre07b}, where $z$ = $d$ sin $b$ is the perpendicular
distance from the Galactic plane and $b$ Galactic
latitude. Integrating Eq.\ref{density} over a sphere with a radius of
200~pc gives $\sim$ 280 as the total number of DNe within 200~pc. If
the space density of DNe follows the space density of CVs as in
\citet{pre07b}, i.e., $\rho_{0}$ = 1.1$^{+2.3}_{-0.7}$ $\times$
10$^{-5}$ pc$^{-3}$, and if a typical DN has an X-ray luminosity
corresponding to the mean luminosity (2 $\times$ 10$^{31}$ erg
s$^{-1}$) of our sample of 11 sources (BZ~UMa excluded), then the
2--10~keV X-ray emissivity of all DNe in the solar neighbourhood would
be 5.5$^{+11.5}_{-3.5}$ $\times$ 10$^{27}$ erg s$^{-1}$
M$_{\odot}^{-1}$ (these account for the uncertainty on the space
density, assuming that this is the dominant source of uncertainty for
the X-ray emissivity of DNe). This would account for more than 100 per
cent of the GRXE emissivity. If DNe were uniformly distributed in the
solar neighbourhood, the X-ray emissivity would be overestimated also
in this case (by 20--30 per cent). However, in both cases, one should
remember that the calculated X-ray emissivity of all DNe within 200~pc
is likely overestimated by the brighter sources in our sample, thus
the calculations give excess emission.

\subsection{Correlations between X-ray luminosity and other parameters}\label{correlation}
\begin{figure}
\centering 
\includegraphics[width=70mm,angle=0]{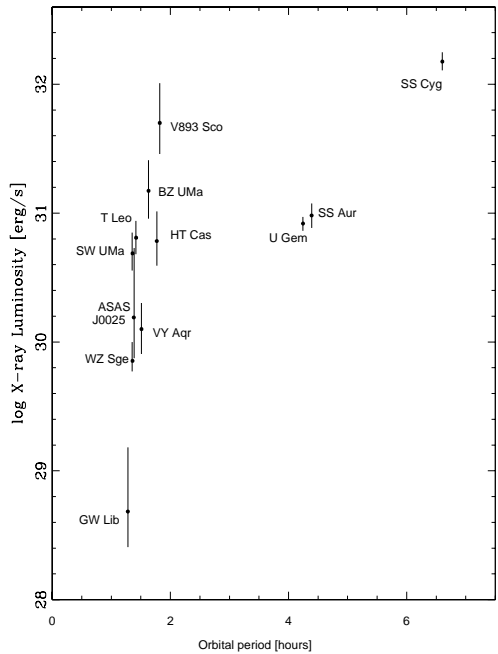}
\caption{The X-ray luminosities (2.0--10.0~keV) versus orbital periods of the source sample.}
\label{orb}
\end{figure}

\begin{figure}
\centering
\includegraphics[width=70mm,angle=0]{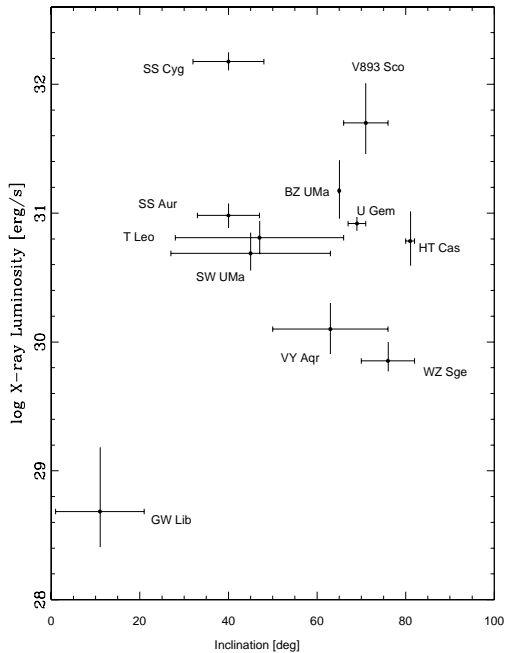}
\caption{The X-ray luminosities (2.0--10.0~keV) versus inclinations of the source sample.}
\label{incl}
\end{figure}

In order to understand whether the X-ray luminosity and the various
parameters (inclination $i$, orbital period $P_{orb}$, shock
temperature $kT_{max}$ and white dwarf mass $M_{WD}$) are correlated,
we carried out Spearman's rank correlation test. Plotting X-ray
luminosity versus a few of these parameters ($i$ and $P_{orb}$) shows
that GW~Lib seems to appear as an outlier compared to the rest of the
sample (Fig.~\ref{orb} and \ref{incl}). Thus, to explore how the
presence/absence of GW~Lib affects the test results, two test cases
were used: 1) GW~Lib was included, and 2) GW~Lib was excluded from the
rest of the sample. In addition, we investigated whether a correlation
between the white dwarf masses $M_{WD}$ and the shock temperatures
$kT_{max}$ (Fig.~\ref{mass_temp}) exists, although in this case,
GW~Lib seems to follow the rest of the sample, thus, carrying out test
case 2) was not necessary.

A strong correlation was found at the 99.95 per cent significance
level (2.8$\sigma$) between the X-ray luminosities and orbital periods
(Fig.~\ref{orb}) when GW~Lib is included in the sample. The
correlation still holds when GW~Lib is excluded (significance is 99.67
per cent). \citet{bas05} noted that there was a weak correlation
between the X-ray luminosities and the orbital periods in their {\it
ASCA} sample, concluding that the X-ray luminosity probably also
correlates with long-term mean accretion rate.

The X-ray luminosity and the inclination $i$ are not correlated in
either case (Fig.~\ref{incl}). The correlation between these
parameters was measured when the inclination of BZ~UMa was set to
65$^{\circ}$, and altering the inclination between 60$^{\circ}$ and
75$^{\circ}$ did not affect the result. Since no correlation was
found, this result is in contrast with the discovery of
anti-correlation between the emission measure and inclination found by
\citet{van96}. It is worth noting that the {\it ROSAT} bandpass was
very narrow, covering only 0.1--2.4 keV where the softer X-ray
emission (and more luminous emission) is probably intrinsically
absorbed by the sources. In addition, an anti-correlation between the
X-ray luminosity and inclination was also seen by \citet{bas05} in the
{\it ASCA} sample, although, Baskill et al. noted that the
inclinations might be uncertain, and this can also be the case in our
sample.

Finally, the white dwarf masses $M_{WD}$ and the shock temperatures
$kT_{max}$ correlate with a significance of 98.5 per cent when the
mass of VY~Aqr is 0.80 M$_{\odot}$, but becomes less significant (97.4
per cent) if the mass is 0.55 M$_{\odot}$. Of the rest of the
parameters, i.e. the X-ray luminosity $L_{x}$ versus $kT_{max}$ and
$M_{WD}$, $kT_{max}$ showed evidence of correlation with $L_{x}$ at a
significance of 97.6 per cent when GW~Lib was included in the sample,
but $L_{x}$ and $M_{WD}$ had a much lower correlation significance (69
per cent) when including GW~Lib. For the latter correlation test
($L_{x}$ versus $M_{WD}$), the result was the same with both $M_{WD}$
values for VY~Aqr. Excluding GW~Lib decreased the significance to 91
per cent ($L_{x}$ versus $kT_{max}$) and to 63 per cent ($L_{x}$
versus $M_{WD}$).

\section{Conclusions}
We have analysed the X-ray spectra of 13 dwarf novae with accurate
parallax-based distance estimates, and derived the most accurate shape
for the X-ray luminosity function of DNe in the 2--10~keV band to date
due to accurate distance measurements and due to the fact that we did
not use an X-ray selected sample.

The derived X-ray luminosities are located between $\sim$
10$^{28}$--10$^{32}$ erg s$^{-1}$, showing a peak at $\sim$ 10$^{30}$
erg s$^{-1}$. Thus, we have obtained peak luminosities which are lower
compared to other previous studies of CV luminosity functions. The
shape of the X-ray luminosity function of the source sample suggests
that the two following scenarios are possible: 1) the sample can be
described by a power law with a single $\alpha$ slope, but the sample
becomes more incomplete below $\sim$ 3 $\times$ 10$^{30}$ erg s$^{-1}$
than it is above this limit, or, 2) the shape of the real X-ray
luminosity function of dwarf novae is a broken power law with a break
at around 3 $\times$ 10$^{30}$ erg s$^{-1}$.

The integrated luminosity between 1 $\times$ 10$^{28}$ erg s$^{-1}$
and the maximum luminosity of the sample, 1.50 $\times$ 10$^{32}$ erg
s$^{-1}$, is 1.48 $\times$ 10$^{32}$ erg s$^{-1}$. In order to better
constrain the integrated luminosity and the slope of the X-ray
luminosity function, more dwarf novae need to be included in the
sample. Thus, we suggest more future X-ray imaging observations of
dwarf novae in the 2--10~keV band with accurate distance
measurements. The total X-ray emissivity of the sample within a radius
of 200~pc is 1.81 $\times$ 10$^{26}$ erg s$^{-1}$ M$^{-1}_{\odot}$
(2--10~keV). This accounts for $\sim$ 16 per cent of the total X-ray
emissivity of CVs as estimated by \citet{saz06}, and $\sim$ 5 per cent
of the Galactic Ridge X-ray emissivity.

The X-ray luminosities and the inclinations of our sample do not show
anti-correlation which has been seen in other previous correlation
studies, but a strong correlation is seen between the X-ray
luminosities and the orbital periods. Also, evidence for a correlation
between the white dwarf masses and the shock temperatures exists. In
the future, larger dwarf nova samples are needed in order to confirm
these results.

\section*{Acknowledgments}
This research has made use of data obtained from the {\it Suzaku}
satellite, a collaborative mission between the space agencies of Japan
(JAXA) and the USA (NASA). JO acknowledges support from STFC. Part of
this work is based on observations obtained with {\it XMM-Newton}, an
ESA science mission with instruments and contributions directly funded
by ESA Member States and the USA (NASA). We thank the reviewer
M. Revnivtsev for his helpful comments on this paper.

\label{lastpage}
\end{document}